\documentclass{article}%
\usepackage[utf8]{inputenc}
\usepackage[english]{babel}
\usepackage{csquotes}
\usepackage{graphicx}
\usepackage{amsmath,amsfonts,amssymb}
\usepackage[normalem]{ulem}
\usepackage{authblk}
\usepackage[parfill]{parskip}
\usepackage{hyperref}
\usepackage{cleveref}
\usepackage{algorithm}
\usepackage{algpseudocode}
\usepackage{bm}
\usepackage{todonotes}
\setuptodonotes{inline}
\usepackage[a4paper, total={6.77777in, 9in}]{geometry}
\usepackage[font=footnotesize]{caption}
\usepackage[style=authoryear, backend=bibtex, bibencoding=utf8, useprefix=true, maxcitenames=7, maxbibnames=25]{biblatex}
\creflabelformat{equation}{#2\textup{#1}#3}
\title{Multihead self-attention in cortico-thalamic circuits}

\author[a,b]{Arno Granier}
\author[a,c,$\ast$]{Walter Senn}

\affil[a]{{\small Department of Physiology, University of Bern, Switzerland}}
\affil[b]{{\small Graduate School for Cellular and Biomedical Sciences, University of Bern, Switzerland}}
\affil[c]{{\small Center for Artificial Intelligence in Medicine, University of Bern, Switzerland}}
\affil[$\ast$]{{\small correspondence: walter.senn@unibe.ch}}

\begin{document}
\maketitle

\begin{abstract}\small
  \noindent Both biological cortico-thalamic networks and artificial transformer networks use canonical computations to perform a wide range of cognitive tasks.
  In this work, we propose that the structure of cortico-thalamic circuits is well suited to realize a computation analogous to multihead self-attention, the main algorithmic innovation of transformer networks.
  We assign distinct computational roles to superficial and deep pyramidal cells of the cortex: while superficial pyramidal cells maintain a key-value memory, deep pyramidal cells encode the current query, gain-modulated by the key-value memory in the superficial layer.
  We show that the structure of this computation matches the fine-grained structure of core and matrix projections from the thalamus to the cortex.
  We then suggest the parallel between one head of attention and a cortical area, and propose that a thalamo-cortico-thalamic pathway implements a computation akin to a multihead, unnormalized, linear self-attention block.
  Cross-attention corresponds to the key-value memory of one cortical area being used for retrieval by the query in another cortical area. 
  Finally, as a first step towards a mechanistic theory of synaptic learning of cortical transformers, we derive the formal gradients of a typical loss function with respect to the parameters of such computation.
\end{abstract}


\section*{Introduction}

Transformer networks \parencite{vaswani2017attention} have taken the world by storm, revolutionizing natural language processing and cognition.
The high performance of this network architecture is not restricted to language, but extends to any domain where inputs can be expressed as sequences of `tokens', including vision \parencite{dosovitskiy2021an} and abstract reasoning \parencite{chollet2024openai}.
While in language processing a token corresponds to a syllable or word, it corresponds to a `foveal patch' in vision, or a `memory item' in abstract reasoning.
At the hearth of transformer networks is the self-attention mechanism, a scalable and parallelizable computation that can handle long-range dependencies when processing sequences. 
Searching for a possible cortical implementation of such self-attention mechanisms may give hints on how cortex may reach its cognitive power.

Self-attention is based on the computation of three different quantities by linear projections of the same sequence elements (tokens): queries, keys, and values.
An intuition for the different roles of these quantities is best given in terms of a soft dictionary: each key indexes a specific dictionary item (a `value'), and upon a query, a mixture of values is returned. 
Each value of this mixture is weighted by the similarity between the query and the key corresponding to that value.
The initial interpretation of self-attention was as a \textit{parallel} computation over all tokens in a temporal context window \parencite{vaswani2017attention}. In this interpretation, current and past tokens each compute their own key, query, and value. Each query is compared to all keys forming a multiplicative attention mask over values, and the resulting weighted sum for each token forms the output.
A formally equivalent formulation (see \nameref{sec:methods}) 
is as a key-value \textit{recurrent} memory that online sums up new key-memory pairs, queried by the current token of the sequence \parencite{katharopoulos2020transformers,sun2023retentive,gershman2025key}. In this interpretation, only the current token computes a key, a query, and a value. The current key-value pair is integrated into a leaky recurrent memory, and the current query retrieves from the recurrent memory a weighted sum of past values based on the similarity between the associated keys in the memory and the new query.
In a multihead self-attention block, this mechanism is repeated in parallel across multiple independent attention heads, each computing different keys, queries, and values, thus attending to different aspects of the tokens. To form the output of a multihead self-attention block, the results of all heads are added together with output weights specific to each head \parencite{elhage2021mathematical}.

The neocortex of mammals is also built on a parallelizable canonical computational motif \parencite{douglas2004neuronal,harris2015neocortical,powell2024common,meyer2025expansion} repeated both in parallel and in series \parencite{felleman1991distributed,d2022hierarchical}.
Concerted efforts have recently greatly advanced the data available on the local components of this canonical cortical circuit \parencite{staiger2021neuronal,sievers2024connectomic,microns2025functional}, but a computational interpretation with demonstrably high performance on hard tasks is missing.
Moreover, there is a growing interest in computational interpretations of the crucial interactions between the cerebral cortex and the thalamus \parencite{suzuki2023deep,furutachi2024cooperative,sherman2024transthalamic,mckinnon2025disruption,whyte2025burst}.

Previous work on biologically-plausible neural implementations of self-attention have focused on a single head, and aim to model either the hippocampal formation \parencite{whittington2022relating} or the role of glial cells \parencite{kozachkov2023building}.
Claims of biological plausibility have also been made on the basis of the equivalence between self-attention and the update in modern continuous Hopfield networks \parencite{ramsauer2021hopfield,krotov2021large,hoover2023energy}. 
Here we propose that the structure of cortico-thalamic circuits, cell types, pathways and interactions, are well suited to implement a computation akin to multihead self-attention.
Specifically, we base ourselves on the formulation of linear self-attention as a recurrent key-value memory system \parencite{katharopoulos2020transformers,sun2023retentive,gershman2025key}.
We show how cortical layer 2/3 pyramidal neurons can form a key-value memory that modulates the query representation in layer 5 pyramidal neurons. 
We do not claim that cortico-thalamic circuits strictly implement the same computation as transformer networks, but rather, that the reported analogies might be illuminating for a mechanistic understanding of cortical computation and, eventually, on the cognitive ability of the cortex.

\section*{Results}

\subsection*{Multihead linear self-attention}
In response to an input token $\bm x_t\in\mathbb{R}^{d_e}$ at time $t$ (a column vector), head-specific keys and values are calculated, $\bm k^h_t = \bm W_K^h \bm x_t$ and $\bm v^h_t = \bm W_V^h \bm x_t$.
Memory matrices $\bm M_t^h \in \mathbb{R}^{d_v\times d_k}$ indexed by `heads' $h$ integrate an association between the current keys and values, alongside those of past key-value memories, 
\begin{equation}
  \bm M_t^h = \gamma\bm M_{t-1}^h + \bm v^h_t \, \phi(\bm k^h_t)^\mathsf{T} \;, \label{eq:M}
\end{equation}
with a discount (leak) factor $\gamma\in[0,1]$ of past memories, and a possible non-linearity $\phi$ applied component-wise to the current key (the subsequent transpose yields a row vector that forms an outer product with the column vector $\bm v^h_t$ yielding the current key-value association). 

The different memory matrices $\bm M_t^h$ can be queried with head-specific queries calculated from the current token,  $\bm q^h_t = \bm W_Q^h \bm x_t$, leading to the retrieved mixture of past values or memory-modulated query representations $\bm M_t^h\,\phi(\bm q^h_t)$.  
An output $\bm y_t$ is constructed by summing the memory-modulated queries across the different heads with head-specific output weights, 
\begin{equation}
  \bm y_t = \sum_h \bm W_O^h \bm M_t^h\,\phi(\bm q^h_t) \;. \label{eq:y}
\end{equation}
\Cref{eq:M,eq:y} describe a multihead, linear, unnormalized self-attention layer (for $\gamma=1$, see \nameref{sec:methods} for more details, and \cite{gershman2025key} for an introduction to the computational concepts). This is a popular alternative to the more classical softmax multihead self-attention, with extensive testing in the recent machine learning literature \parencite{schlag2021linear,qin2022devil,fu2023hungry,von2023transformers,sun2023retentive,yang2024parallelizing}. 

\subsection*{A distinct role for pyramidal cells of cortical layers 2/3 and 5}

\begin{figure}[!ht]
    \centering
    \includegraphics{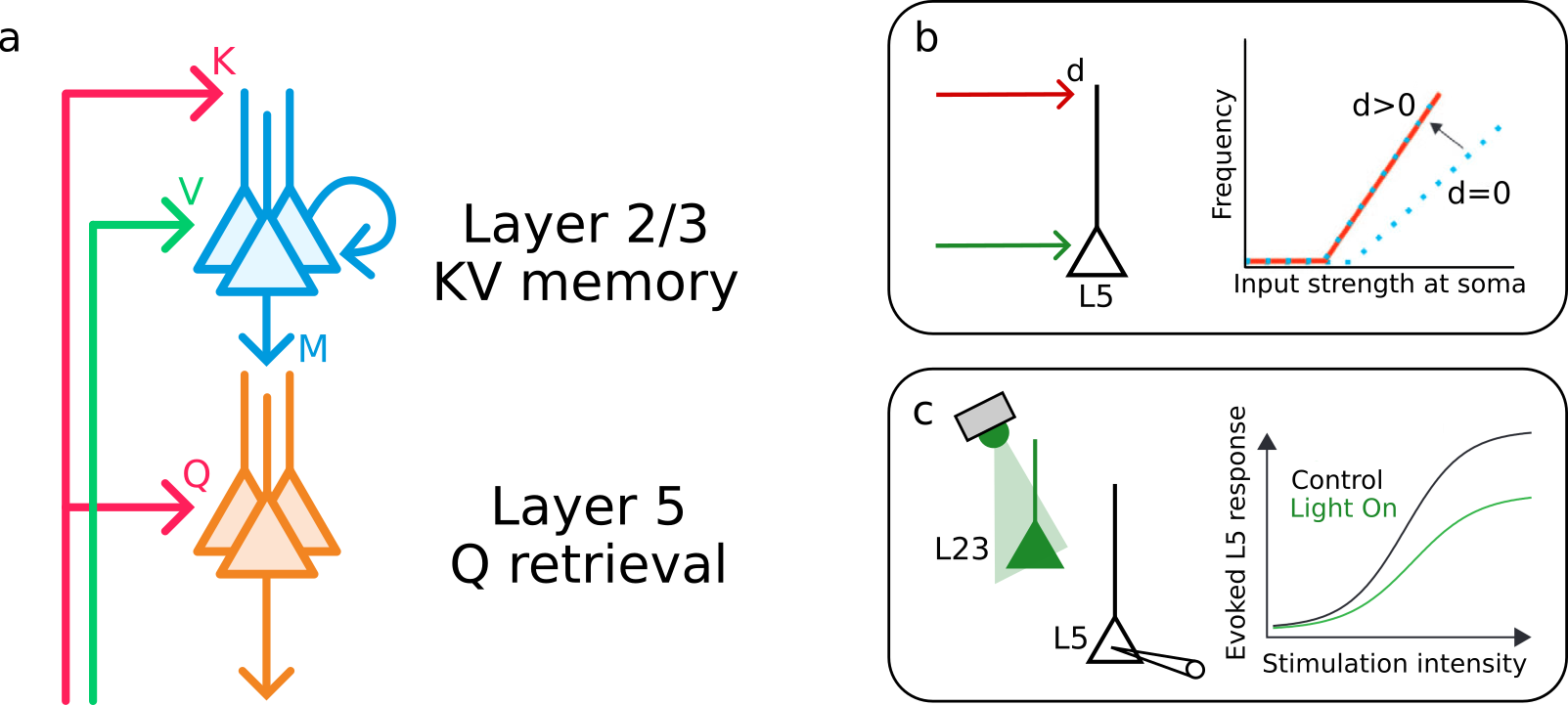}
    \caption{\footnotesize A cortical microcolumn. (a) A layer 2/3 pyramidal ensemble (blue) integrates over time pairs of keys (K, red) and values (V, blue). A layer 5 pyramidal ensemble (orange) retrieves a mixture of past values from the layer 2/3 ensemble (M, blue) based on the current query (Q, red). The layer 5 ensemble sends the output (orange) of the microcolumn. (b) Apical input has a multiplicative impact on the somatic firing of layer 5 pyramidal cells. Adapted from \textcite{larkum2004top}. (b) Layer 2/3 pyramidal cells modulate the gain of layer 5 pyramidal cells. In this experiment layer 2/3 pyramidal cells are photo-inhibited (green line). Adapted from \textcite{quiquempoix2018layer}.
    }
    \label{fig:235}
\end{figure}

To map \cref{eq:M,eq:y} to neural circuits, we first focus on a single head and its internal computation.
We propose that the time-dependent memory matrix $\bm M_t^h$ is encoded by a population of $d_vd_k$ pyramidal ensembles, which we situate in cortical layer 2/3.
A pyramidal ensemble as defined here is a tightly coupled group of pyramidal cells encoding \textit{one} scalar quantity. 
Every layer 2/3 pyramidal ensemble is then tasked with integrating over time the product of a specific component of the value vectors with a specific component of the key vectors (see \cref{eq:M}).
The values $\bm v_h^t$ are postulated to be represented by basal dendritic input to the layer 2/3 pyramidal neurons, while the keys $\bm k_h^t$ are represented by the apical dendritic input (\cref{fig:235}\,a). 
The somatic input of such a layer 2/3 pyramidal neuron is formed by the product of the (putatively nonlinear) apical activity times the basal activity. The individual layer 2/3 pyramidal neurons hence comprise the components of the $d_v\!\times\!d_k$ matrix $\bm v^h_t \, \phi(\bm k^h_t)^\mathsf{T}$.
The integration with past activity, yielding $\bm M_t^h$, is enabled by the recurrent connections forming a short-term memory of the processed stimuli \parencite{Holmgren2003,Funayama2015}. 

To each layer 2/3 ensemble (a component of $\bm M_t^h$) we attach one ensemble of layer 5 pyramidal neurons, forming a cortical microcolumn or unit module, see \cref{fig:235}\,a. 
The basal dendritic inputs to the layer 5 pyramidal ensembles represent the query $\bm q_t^h$, while the apical dendritic inputs to layer 5 represent a component of the key-value memory $\bm M_t^h$.   
Every layer 5 ensemble is tasked with computing the product of its associated layer 2/3 ensemble and the corresponding component of the query vector. 
Projections from layer 5 ensembles, e.g.~to a higher thalamic kernel, form the output $\bm y_t$ of the head (see \cref{eq:y,eq:wi}).

To reiterate: layer 2/3 pyramidal cells store key-value pairs over time, while layer 5 pyramidal cells represent the mixture of values retrieved by the current query.
Both operations demand that pyamidal ensembles compute the product of their two classes of dendritic inputs, as has been shown in particular for layer 5 pyramidal cells (see \cref{fig:235}\,b; \cite{larkum2004top}).
The local pathway from layer 2/3 to layer 5 pyramidal cells is established as a major component of the classical local canonical cortical circuit \parencite{douglas2004neuronal}. 
Moreover, as demanded by our interpretation, there is experimental evidence that layer 2/3 pyramidal cells act as controllers of the gain of layer 5 pyramidal cells (see \cref{fig:235}\,c; \cite{quiquempoix2018layer}).

\subsection*{Core and matrix thalamo-cortical projections compute keys, queries, and values}

\begin{figure}[!ht]
    \centering
    \includegraphics{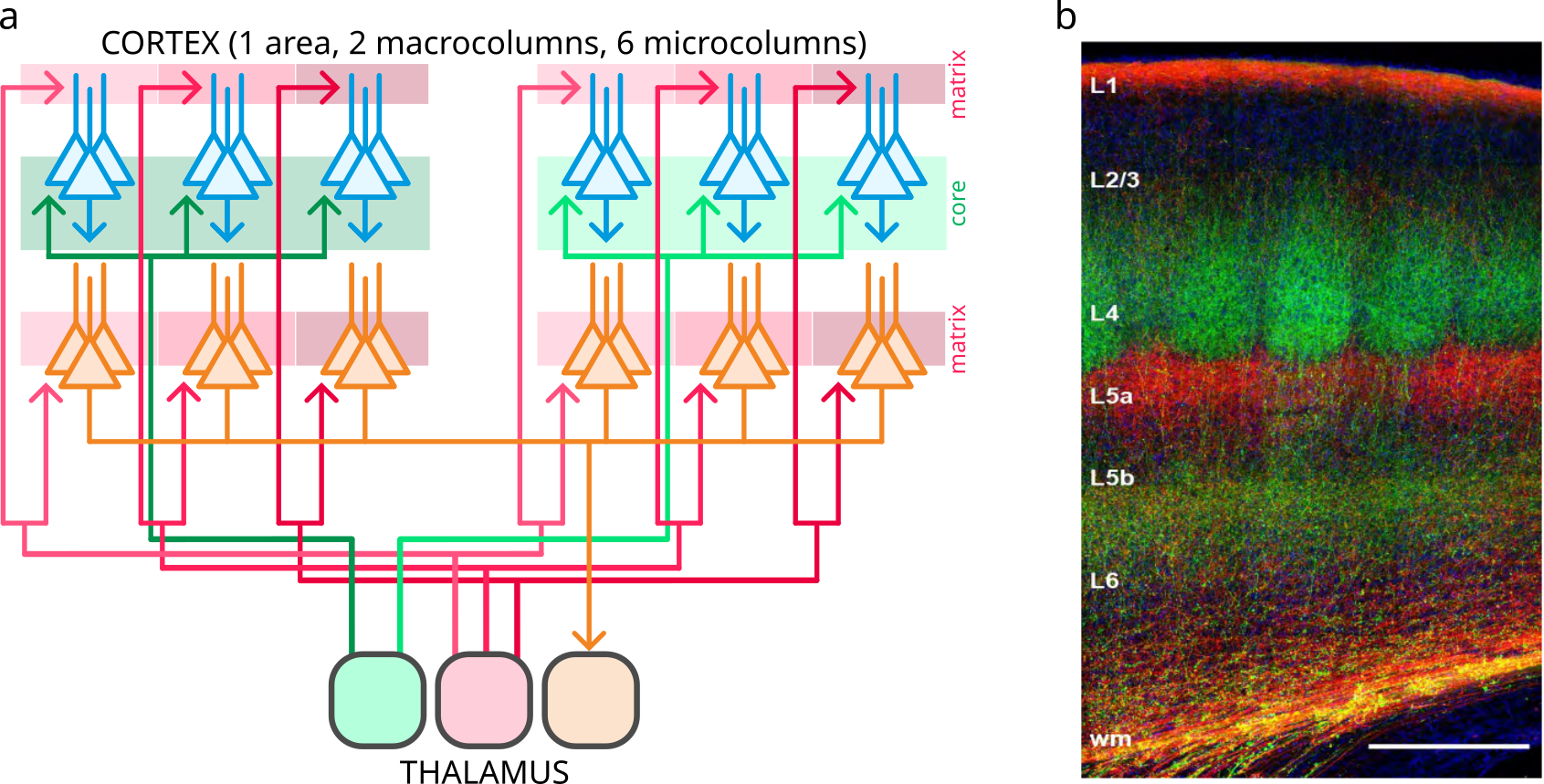}
    \caption{\footnotesize Core/Matrix structure of thalamo-cortical projections may compute keys, queries, and values.
    (a) One head of linear self-attention with cortical microcolumns ($d_v=3$, $d_k=2$).
      Green and red bottom squares represent a thalamic nucleus encoding the current input token $\bm x_t$ and send respectively core-type and matrix-type projections to the cortex.
      Shades of colours indicate which component of the key, query, and value vector a particular ensemble is receiving.
      One vertical set of a layer 2/3 and layer 5 pyramidal ensemble act as in \cref{fig:235}\,.
      Orange projections contact a (higher-order) thalamic nucleus represented by an orange square.
    (b) Fluorescence imaging data for core (green) and matrix (red) thalamo-cortical projections reproduced from \textcite{staiger2021neuronal}. Scale bar is 250\textmu m.
    }
    \label{fig:thal}
\end{figure}

To realize \cref{eq:M,eq:y} using $d_vd_k$ cortical microcolumn (see \cref{fig:235}), the inputs to these microcolumns need to be specifically distributed.
In a $d_v\times d_k$ grid of microcolumns, value projections ($\bm W_V^h$) need to span horizontally or row-wise, while key and query projections ($\bm W_K^h$, $\bm W_Q^h$) need to span vertically or column-wise.
We opt for a computationally equivalent layout, and split our set of microcolumns into $d_k$ macrocolumns, each comprising $d_v$ microcolumns.
Each microcolumn within one macrocolumn receives the same value ($\bm v^h_t$, green in \cref{fig:thal}\,a), and a different key and query ($\bm k^h_t$ and $\bm q^h_t$, red in \cref{fig:thal}\,a). 
Therefore, value projections need to be dense (contact all pyramidal ensembles in a macrocolumn), focused (project to only one macrocolumn), and target layer 2/3. On the contrary, key and query projections need to be sparse (contact only one ensemble in a macrocolumn), diffuse (project to all macrocolumns), and target layer 2/3 and layer 5 respectively.

This corresponds well to the structure of core and matrix thalamo-cortical projections (\cite{jones1998core,staiger2021neuronal}; but see \cite{sherman2024reconsideration}).
Core projections (green in \cref{fig:thal}\,b) are dense topographic projections targeting mainly cortical layers 3 and 4, and are suggested to convey values ($\bm v^h_t$). They contact the basal dendritic tree of layer 2/3 pyramidal cells either directly or relaying through layer 4 neurons, again a major component of the classical cortical microcircuit \parencite{douglas2004neuronal}.
In at least the well studied rat somatosensory cortex, the extent of core projections in layer 4 are the basis for the definition of macrocolumns or `barrels' \parencite{staiger2021neuronal}.
Matrix projections (red in \cref{fig:thal}\,b) are more diffuse thalamo-cortical projections targeting cortical layers 1 and 5, and are suggested to convey keys and queries ($\bm k^h_t$ and $\bm q^h_t$). In this interpretation, the primary thalamic nuclei encode the original sequence elements or tokens ($\bm x_t$ in \cref{eq:M,eq:y}), while higher thalamic nuclei receive back the key-value modulated queries from all the heads ($\bm y_t$).

\subsection*{Cortical areas as attention heads}

\begin{figure}[!ht]
    \centering
    \includegraphics{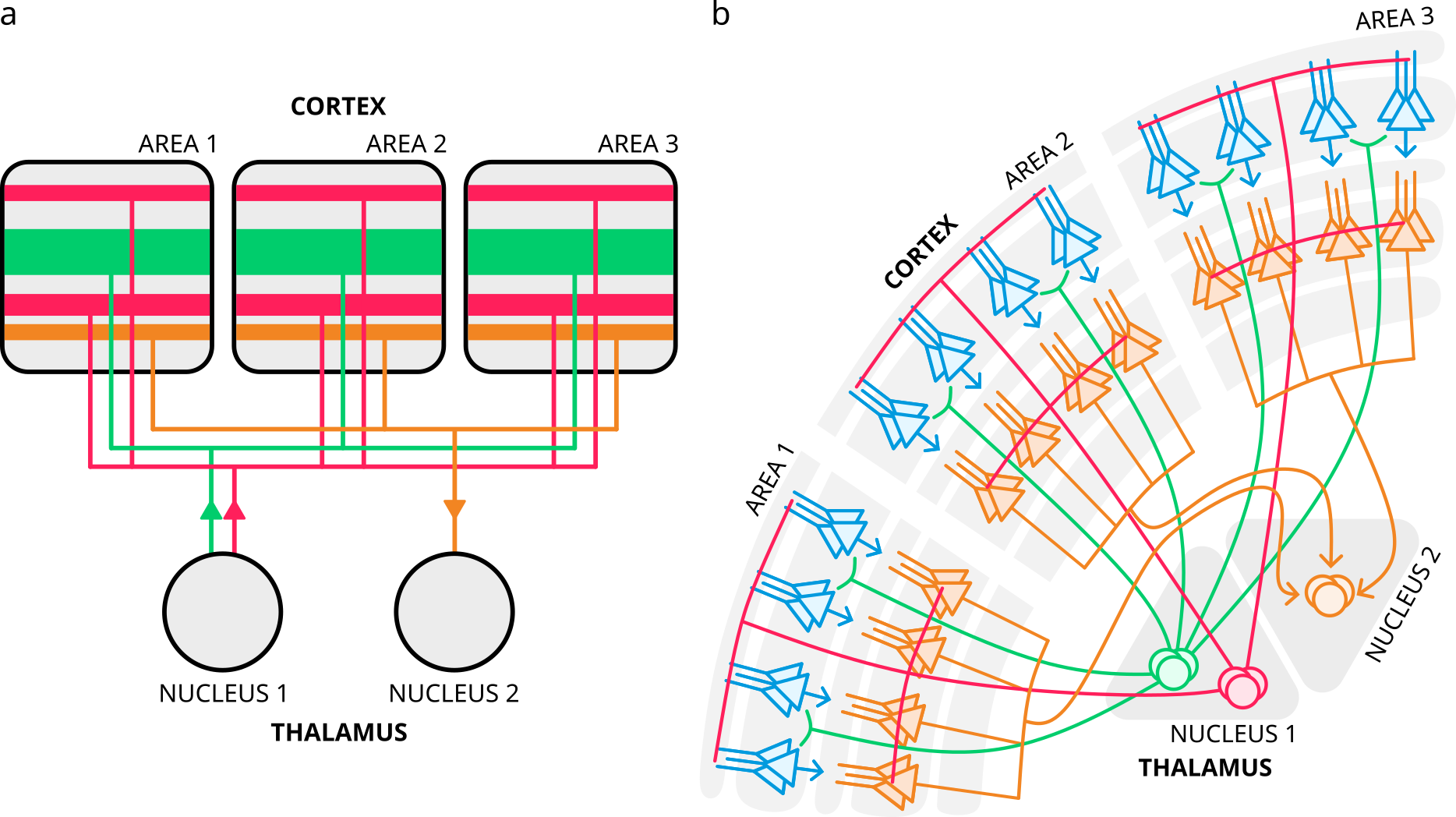}
    \caption{\footnotesize Cortical areas as attention heads. `Transcortical' divergent and convergent pathways for multihead attention. (a) Red (matrix) thalamo-cortical projections compute keys ($\bm k_t^h$, layer 1 component) and queries ($\bm q_t^h$, layer 5 component). Green (core) thalamo-cortical projections compute values ($\bm v_t^h$). The detailed connectivity and computation of one cortical area (square) is depicted in \cref{fig:thal}. Orange cortico-thalamic projections from layer 5 pyramidal cells form the output of an area/head, the key-value modulated queries $\bm M_t^h\,\phi(\bm q^h_t)$,  and are summed up across areas/heads ($h)$ in a higher-order thalamic nucleus. 
 (b) A more biological depiction of the same concepts. }
    \label{fig:area}
\end{figure}

We have until now described the computation in a single attention head of a transformer network. 
Multihead self-attention entails running multiple such heads independently in parallel and integrating their results additively with head-specific output weights (see \cref{eq:y}\,; \cite{elhage2021mathematical}).
We next describe the analogy between one head of attention and one cortical area, see \cref{fig:area}\,.
In doing so, we depart from classical models inspired by machine learning, and do not consider a strict hierarchy of cortical areas (following e.g.~\cite{suzuki2023deep}); rather, at each stage of computation, a subset of cortical areas processes different aspects (`heads') of the same input in parallel (for example in the visual system: shape, color, motion, etc.).
One thalamic nucleus projects to multiple cortical areas in parallel, computing independently in each of them different keys, queries, and values from the same current sequence element. 
The output of each area is integrated additively in another, higher-order thalamic nucleus to which all layer 5 pyramidal cells in the relevant cortical areas project. 
The activity of this higher-order thalamic nucleus is then the input of the next stage of processing, forming the cortical hierarchy in the classical sense.

In other words, the cortico-thalamic complex forms a shallow hierarchy where cortical areas process different aspects of the input in parallel before integrating the results back in the thalamus \parencite{mumford1991computational}.
This general organization into `transcortical' pathways connecting two thalamic nuclei through a subset of cortical areas is consistent with the observation that single thalamic cells integrate information from different cortical areas \parencite{sampathkumar2021integration}, and that layer 5 pyramidal cells avoid driving back the thalamic nuclei projecting to their cortical area \parencite{cassidy2025complementary}.

\subsection*{Thalamic and cortical cross-attention}
A natural extension of our model is to include cross- (rather than self-) attention (e.g.~as already present in \cite{vaswani2017attention}). In that case, the current token $\bm z_t$ \textit{from another sequence} $\{\bm z_1, \dots, \bm z_t\}$ queries the key-value memory formed by the first sequence $\{\bm x_1, \dots, \bm x_t\}$.
The two different sequences may encode information from two different modalities. For example, the tokens $\bm x_t$ may encode auditory input and $\bm z_t$ visual input in the corresponding primary thalamic kernel. 

\begin{figure}[!ht]
    \centering
    \includegraphics{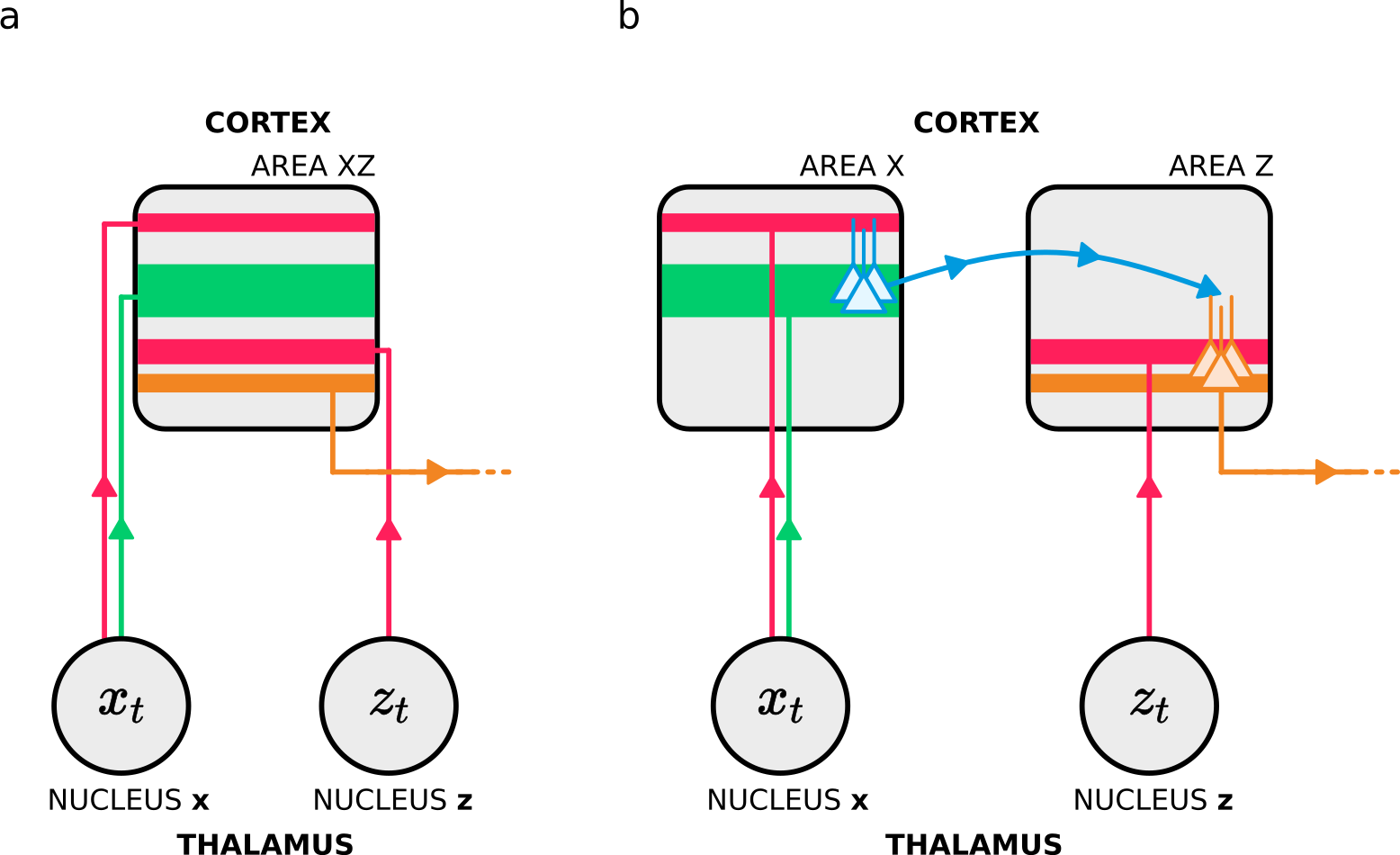}
    \caption{\footnotesize Two circuits for cross-attention. 
    (a) A purely thalamo-cortical implementation involving an associative cortical area. Two thalamic nuclei encode the current element of two different sequences $\bm x_t$ and $\bm z_t$ and project to a single cortical area. Projections from the nucleus encoding $\bm x_t$ target the superficial cortical layers and update the key-value memory. Projections from the nucleus encoding $\bm z_t$ target cortical layer 5 and query the memory.
    (b) An implementation with interareal cortico-cortical projections. Each thalamic nucleus encoding the current element for different sequences project to a different cortical area. Nucleus $\bf x$ updates the key-value memory in layer 2/3 of area $X$. Crucially, cortical area $X$ sends to cortical area $Z$ its key-value memory via an interareal projection (blue). In area $Z$, the key-value memory of area $X$ is integrated with the query from thalamic nucleus $\bf z$ in layer 5 pyramidal cells (orange), and form the output of that second area $Z$ to another thalamic nucleus or another cortical area. }
    \label{fig:ca}
\end{figure}

Structurally, cross-attention could be realized in multiple ways.
A first realization entails thalamo-cortical projections from the two different thalamic kernels to the same associative cortical area (see e.g.~\cite{Cappe2009}, \cref{fig:ca}\,a). 
The auditory thalamic kernel encoding $\bm x_t$ targets layer 2/3 via core projections, and layer 1 via matrix projections.
It updates the key-value memory $\bm M_t^{XZ}$ in the layer 2/3 pyramidal neurons of the associative area $XZ$ as in eq.~\ref{eq:M},
\begin{equation}
\bm M_t^{XZ} = \gamma\bm M_{t-1}^{XZ}+\bm W_V^{XZ} \bm x_t \,\phi(\bm W_K^{XZ} \bm x_t)^\mathsf{T}
\;. 
\end{equation}

Queries are calculated based on the inputs $\bm z_t$ from the visual thalamic kernel to layer 5 of the same area $XZ$ via matrix projections, $\bm W_Q^{XZ} \bm z_t$. The output to an associative thalamic nucleus is calculated from the memory-modulated query representation,
\begin{equation}
  \bm y_t^{XZ} = \bm W_O^{XZ} \, \bm M_t^{XZ} \, \phi(\bm W_Q^{XZ} \bm z_t) \;. 
    \label{eq:cross1}
\end{equation}

A second realization of cross-attention entails cortico-cortical projections between two different cortical areas $X$ and $Z$, say the primary auditory and visual cortices, driven by different thalamic kernels (see e.g.~\cite{Ibrahim2016}, \cref{fig:ca}\,b).
In area $X$ the key-value memory $\bm M_t^{X}$ is calculated in layer 2/3 based on input $\bm x_t$ from the auditory kernel, say. This memory projects to the apical dendrites of layer 5 pyramidal neurons in area Z. The basal dendrites of these layer 5 pyramidal neurons calculate the queries $\bm W_Q^Z$ based on the input $\bm z_t$ from the visual kernel.
These queries may then be modulated by the memory $\bm M_t^{X}$ from area $X$, in addition to a possible memory modulation by $\bm M_t^{Z}$ from within the same area $Z$,
\begin{equation}
  \bm y_t^{X,Z} = \bm W_O^Z \, (\bm M_t^X +  \bm M_t^Z ) \, \phi(\bm W_Q^Z \bm z_t) \;. 
  \label{eq:cross2}
\end{equation}
The output $\bm y_t$ may project back to yet another thalamic kernel (or another cortical area, \cref{fig:ca}\,b).

The original transformer version in \cite{vaswani2017attention} considers either self-attention ($\bm M_t^Z$) or cross-attention ($\bm M_t^X$ in eq.~\ref{eq:cross2}, also shown in \cref{fig:ca}\,b) only. Yet, the combination of self- and cross-attention, as expressed in eq.~\ref{eq:cross2}, seems to be an ubiquitous motif of associative and sensory cortical areas. 
In fact, attentional modulation via top-down input to apical dendrites of cortical pyramidal neurons, implementing cross-attention, has been shown to be involved in perceptional tasks in mice and monkeys \parencite{Manita2015,Nandy2017}. 
Apical and basal dendrites of cortical pyramidal neurons may be simultaneously innervated from either the same cortical area \parencite{Petreanu2009} or the same thalamic kernel \parencite{Rah2013,Balcioglu2023}, implementing self-attention.

\subsection*{Synaptic plasticity}
A full cortical implementation of recurrent transformers entails synaptic plasticity rules that follow the gradients of the loss function by which the transformers are trained. 
We consider an energy-based description of transformer networks from which we formally derive gradient-based plasticity rules for all the involved synapses ($\bm W_O^h$, $\bm W_V^h$, $\bm W_K^h$, and $\bm W_Q^h$, \nameref{sec:methods}, \cref{eq:dWo,eq:dWv,eq:dWk,eq:dWq}). 
For a linear transfer function $\phi$, these learning rules can be written as a product of local (in space and time) quantities multiplied by slow synaptic variables (\nameref{sec:methods}, \cref{eq:dWvr,eq:dWkr,eq:dWqr,eq:Sv,eq:Sk,eq:Sq}). These rules could potentially be locally implemented by dedicated dendritic compartments that sum appropriately modulated backpropagated errors (\nameref{sec:methods}, \cref{eq:dotwV,eq:dotwK,eq:dotwQ}), following ideas from \parencite{Sacramento2018,senn2024neuronal}. 

\section*{Discussion}
\textbf{Summary.} In this work, we highlighted analogies between the structure of cortico-thalamic circuits and multihead, unnormalized, linear self-attention.
In particular, we proposed that the pyramidal cells of layer 2/3 implement an associative key-value memory \parencite{gershman2025key}, while the pyramidal cells of layer 5 represent the queries that are multiplicatively modulated by the L2/3 memories.
The fine-grained structure of core and matrix projections from the thalamus to the cortex, and the cortical gain modulation, is well suited to implement the computation underlying the memory- or attention-modulated query representation.
Different cortical areas can be interpreted as different heads of attention in a transformer block that are integrated e.g.~in a higher thalamic nucleus.
Finally, we interpret cross- (rather than self-) attention as a particularly important computation for associative cortices, and discussed potential circuit implementations.

\textbf{From multihead self-attention to a full transformer block.} Reproducing the full computation of a transformer block entails the sequential computation of multihead self-attention and a fully-connected network with one hidden layer, together with normalization. 
In this work, we have covered the central innovation of transformer networks, multihead self- and cross-attention.
It is left for future work to analyze whether analogs of the remaining computation of the hidden layer could be realized by multi-synaptic cortico-cortical projections, multi-synaptic thalamo-cortical loops via pyramidal cells of cortical layer 6, local circuits, or projections via cortical layer 4.
The remaining normalization is well-studied in neuroscience and has been described as a canonical neural computation in terms of divisive normalization \parencite{carandini2012normalization,shen2021correspondence} or soft winner-take-all over dendritic activities \parencite{Murayama2009}. 

\textbf{Parallel processing of multiple tokens.} There exist alternative mappings of multihead self-attention to cortico-thalamic circuits. 
Earlier versions of this work considered the computation of an attention signal in layer 2/3 pyramidal cells by matching multiple keys to queries in parallel, and the computation of attention-modulated values in layer 5 pyramidal cells by modulating values with this attention signal.
This was inspired by the parallel rather than recurrent formulation of self-attention.
The main drawback is the need to encode the full temporal context, leading to a quadratic scaling in context length, avoided by the recurrent formulation. 
Other works consider biologically plausible encoding of a temporal context based on cortical traveling waves \parencite{muller2024transformers}.

\textbf{Cross-attention from hippocampal memory.} In this work we have proposed that layer 2/3 pyramidal cells encode a short-term `working' memory of the immediate history by integrating key-value pairings.
Mechanisms by which tokens outside a temporal context window are discarded or forgotten, or more elaborate selective forgetting mechanisms \parencite{schlag2021linear,yang2024parallelizing}, could yet be implemented. 
Interestingly, long-term memory might also rely on a key-value memory system, likely involving the hippocampal and parahippocampal structures \parencite{whittington2022relating,gershman2025key} and their projection to cortical layer 1 \parencite{doron2020perirhinal}. These hippocampal memory projections could target the apical layer-1 tuft dendrites of layer 5 pyramidal cells (that are also targeted by local layer 2/3 pyramidal cells, see \cite{ledderose2023layer}).


\textbf{Cortico-cortical interareal projections.} We have suggested a potential role of interareal cortico-cortical projections in realizing a form of cross-attention. 
More generally, the structure of cortico-cortical feedforward and feedback connections respectively resembles the structure of core and matrix projections in their laminar targets \parencite{markov2014anatomy, harris2019hierarchical}.
The computation realized by these projections might then be analyzed in our framework as participating in the formation of values (core, feedforward), and keys and queries (matrix, feedback).
In support, recent evidence from direct inactivation of cortico-cortical feedback projections demonstrates their role in attentional gain modulation \parencite{debes2023suppressing}.

\textbf{Layer 6b and perceptual attention.} Finally, excitatory neurons in cortical layer 6b send local projections to the same layers that matrix thalamo-cortical projections target \parencite{zolnik2024layer} and have been hypothesized to underlie attentional modulation in perception \parencite{zolnik2025layer}. Given that, in our work, matrix thalamo-cortical projections compute keys and queries also underlying attention, an extension of our model to include layer 6b, and distinguish the notions of self-, cross- and perceptual attention, seems promising.


\section*{Methods} \label{sec:methods}

\subsection*{Linear self-attention}
Let $\bm x_1, \dots, \bm x_T$ be a sequence of $T$ elements or tokens, with $\bm x_t \in \mathbb{R}^{d_e}$. The self-attention mechanism central in transformer networks relies on three linear projections of sequence elements
\begin{equation}
  \bm q_t = \bm W_Q \bm x_t~~;~~\bm k_t = \bm W_K \bm x_t~~;~~\bm v_t = \bm W_V \bm x_t \;,
\end{equation}
with $\bm q_t \in \mathbb{R}^{d_k}$, $\bm k_t \in \mathbb{R}^{d_k}$, $\bm v_t \in \mathbb{R}^{d_v}$, $\bm W_Q \in \mathbb{R}^{d_k \times d_e}$, $\bm W_K \in \mathbb{R}^{d_k \times d_e}$, and $\bm W_V \in \mathbb{R}^{d_v \times d_e}$.

We consider computation with causal masking, that is, where a token at time $t$ can only attend to other tokens at times $p\leq t$ in the past.
A classical self-attention mechanism (e.g.~\cite{vaswani2017attention}) uses the softmax or normalized exponential dot product as a similarity measure and outputs
\begin{equation}\label{eq:softmaxSA}
  \bm o_t = \sum_{p\leq t} \frac{\exp(\bm k_{p}^\mathsf{T}\bm q_t)}{\sum_{p\leq t} \exp(\bm k_i^\mathsf{T}\bm q_t)}\bm v_{p} \;.
\end{equation}
In this work, we rather use the alternative \textit{linear} self-attention formulation, replacing the exponential kernel with a linear product of feature maps $\phi:\mathbb{R}^{d_k} \rightarrow \mathbb{R}^{d_f}$ \parencite{katharopoulos2020transformers,schlag2021linear}. Multiple work have shown how to approximate \cref{eq:softmaxSA} arbitrarily closely using projections to random features \parencite{peng2021random,choromanski2021rethinking}. Following recent work on linear self-attention, we additionally drop the normalization term \parencite{schlag2021linear,qin2022devil,fu2023hungry,von2023transformers,sun2023retentive,yang2024parallelizing}. This results in outputs, 
\begin{equation}\label{eq:linearSA}
  \bm o_t = \sum_{p\leq t} \phi(\bm k_{p})^\mathsf{T}\phi(\bm q_t)  \,\bm v_{p} \;.
\end{equation}

One advantage of \cref{eq:linearSA} over \cref{eq:softmaxSA} is the possible rearrangement of terms into
\begin{equation}\label{eq:rearrangement}
  \bm o_t 
  = \sum_{p\leq t} \gamma^{t-p} \phi(\bm k_{p})^\mathsf{T}\phi(\bm q_t)\bm v_{p}  
  = \left[\sum_{p\leq t} \gamma^{t-p} \bm v_{p}\phi(\bm k_{p})^\mathsf{T}\right]\phi(\bm q_t)
  = \bm M_t \, \phi(\bm q_t)\;.
\end{equation}
In the last equation, we note $\bm M_t = \sum_{p\leq t} \gamma^{t-p}\bm v_{p}\phi(\bm k_{p})^\mathsf{T}$ the key-value memory, of which a recurrent formulation is \cref{eq:M}. We introduce the forgetting factor $\gamma$ that naturally appears in the recurrent formulation ($\gamma=1$ for an exact match to self-attention). 
The classical formulation of \cref{eq:softmaxSA} bears a computational cost that scales quadratically in time (sequence length). In contrast, the rearrangement of \cref{eq:rearrangement} only scales linearly in time, since each key-value pairing (outer product) can be computed only once, stored (in a key-value memory), and reused for every query \parencite{katharopoulos2020transformers}.

For $\gamma=1$ this infinite recurrent memory is formally equivalent to the parallel formulation with an infinite context length. However, in practice, for long sequences, the parallel formulation considers a fixed context window ($\max(0,t-C) \leq p \leq t$), and earlier tokens dropping out of the window are ignored. 
Following \textcite{sun2023retentive}, we rather consider the case of recurrent `retentive' networks with $\gamma<1$.
We note that more complex forgetting mechanisms have also been proposed, such as a delta rule \parencite{schlag2021linear,yang2024parallelizing}, 
\begin{equation}
  \bm M_t = \bm M_{t-1} - \beta_t(\bm M_{t-1}\phi(\bm k_t) - \bm v_t)\phi(\bm k_t)^\mathsf{T}\;.
\end{equation}


\subsection*{Multihead linear self-attention}
The preceding subsection describes the computation in a single head of linear self-attention. We now extend to $n_h$ heads. The classical presentation of multihead self-attention proposes to concatenate the outputs $\bm o_t^h$ of the different heads and multiplies the resulting vector by output weight $\bm W_O \in \mathbb{R}^{d_e \times n_hd_v}$, yielding the output 
\begin{equation}\label{eq:classicalheads}
  \bm y_t = \bm W_O \left[\bm o_t^1 | \dots | \bm o_t^{n_h}\right]^\mathsf{T} \;.
\end{equation}
An alternative presentation, formally equivalent to \cref{eq:classicalheads} \parencite{elhage2021mathematical}, highlights attention heads as independent and additively integrated with head-specific output weights $\bm W_O^h \in \mathbb{R}^{d_e\times d_v}$,
\begin{equation}\label{eq:heads}
  \bm y_t = \sum_h \bm W_O^h \bm o_t^h \;.
\end{equation}

For our purpose, we remark that the output $\bm y_t$ in \cref{eq:heads} can be written as a sum of gain-modulated layer 5 pyramidal neurons, weighted by the output synapses, 
\begin{equation}\label{eq:wi}
  y_{t,i} = \sum_h \sum_{k} W_{O,ik}^h o_{t,k}^h 
  = \sum_h \sum_{k,j} W_{O,ik}^h  \Big(M_{t,kj}^h \cdot \phi(q_{t,j}^h)\Big) \;.
\end{equation}

\subsection*{Formal gradients of linear self-attention parameters}
Automatic differentiation is now ubiquitous in deep learning, such that formal gradients of the parameters with respect to the loss are rarely derived. However, for a mechanistic theory of gradient-based synaptic learning in our proposed framework, the form of these gradients is central \parencite{Richards2019,richards2023study}. We derive partial derivatives of multihead self-attention with respect to $\bm W_O^h$, $\bm W_V^h$, $\bm W_K^h$, and $\bm W_Q^h$ of a mean squared error loss in a self-supervised, autoregressive setting, that is, when the output $\bm y_t$ should prospectively predict the next input $\bm x_{t+1}$, 
\begin{equation}
  E_t  = \tfrac12\|\bm x_{t+1}-\bm y_t\|^2 = \tfrac12\|\bm e_t\|^2 \;.
  \label{eq:E}
\end{equation}
with $\bm y_t$ defined by \cref{eq:M,eq:y} with $\gamma=1$.
Such an energy function based on prospective prediction errors was recently suggested in the form of a least action principle for cortical computation \parencite{senn2024neuronal}, along with a dendritic implementation of the emerging learning rules \parencite{urbanczik2014learning,Sacramento2018}.

The partial derivatives are
\begin{align}
  -\partial_{\bm W_O^h} E_t &= \bm e_t{\bm o_t^{h\mathsf{T}}} \;, \label{eq:dWo}\\
  -\partial_{\bm W_V^h} E_t &= \sum_{p\leq t} \left[\phi(\bm k_p^h)^\mathsf{T}\phi(\bm q_t^h)\right] \bm b_t^h\bm x_p^{\mathsf{T}} \;, \label{eq:dWv}\\
  -\partial_{\bm W_K^h} E_t &= \sum_{p\leq t} \left[\bm v_p^{h\mathsf{T}}\bm b_t^h\right] \widetilde{\bm q}_{t,p}^h \bm x_p^{\mathsf{T}} \;, \label{eq:dWk}\\
  -\partial_{\bm W_Q^h} E_t &= \sum_{p\leq t} \left[\bm v_p^{h\mathsf{T}}\bm b_t^h\right] \widetilde{\bm k}_{p,t}^h\bm x_t^{\mathsf{T}} \;, \label{eq:dWq}
\end{align}
with $\bm b_t^h = {\bm W_O^{h\mathsf{T}}}\bm e_t$ a backpropagated error, $\widetilde{\bm q}_{t,p}^h=\phi'(\bm k_p^h)\odot\phi(\bm q_t^h)$, and $\widetilde{\bm k}_{p,t}^h=\phi'(\bm q_t^h)\odot\phi(\bm k_p^h)$. We check these results against numerical approximations based on finite differences and forward automatic differentiation\footnote{\url{https://github.com/arnogranier/Formal-MHSA-gradient-numerics}}. These derivatives consist only of outer products, scaled in \cref{eq:dWv,eq:dWk,eq:dWq} by scalar dot products. That is, they consist of products of pre- and post- synaptic quantities, respecting at least a weak notion of locality. Typically, gradient-based plasticity considers synaptic learning rules of the form $\dot{\bm W}_X^h = -\partial_{\bm W_X^h} E_t$. Full circuit implementations are left for future work, but it might be useful to consider that the postsynaptic components of \cref{eq:dWv,eq:dWk,eq:dWq} are encoded in apical dendrites \parencite{urbanczik2014learning}. An extension to the energy $E = \sum_t E_t$ is straightforward.

The temporal summations over past tokens indicate slow variables influencing synaptic plasticity, stored and incremented with the presentation of each new token. In particular, in the case $\phi(x)=x$ (such as in \cite{sun2023retentive,von2023transformers}, etc.), we can write \cref{eq:dWv,eq:dWk,eq:dWq} in a recurrent form, 
\begin{align}
  -\partial_{\bm W_V^h} E_t &= \bm b_t^h\bm q_t^{h\mathsf{T}} \bm S_{V,t}^h \;, \label{eq:dWvr} \\
  -\partial_{\bm W_K^h} E_t &= \bm q_t^h \bm b_t^{h\mathsf{T}} \bm S_{K,t}^h \;, \label{eq:dWkr}\\
  -\partial_{\bm W_Q^h} E_t &=  \bm S_{Q,t}^h \bm b_t^h\bm x_t^{\mathsf{T}}\;,  \label{eq:dWqr}
\end{align}
with matrix-valued slow synaptic variables $S_{V,K,Q}^h$ being recurrently updated  according to
\begin{align}
  \bm S_{V,t}^h &= \bm S_{V,t-1}^h + \bm k_t^h \bm x_t^\mathsf{T} \;, \label{eq:Sv}\\
  \bm S_{K,t}^h &= \bm S_{K,t-1}^h + \bm v_t^h \bm x_t^\mathsf{T} \;, \label{eq:Sk}\\
  \bm S_{Q,t}^h &= \bm S_{Q,t-1}^h + \bm k_t^h \bm v_t^{h\mathsf{T}} \;. \label{eq:Sq}
\end{align}
In this simple case of a linear $\phi$, it is in fact sufficient to keep a single memory variable for both inference and learning integrating $\bm x_t \bm x_t^\mathsf{T}$ as illustrated in \cref{alg:single}. In the more general case where $\phi$ is not identity, inference easily adapts as covered above, but it remains to be investigated whether gradient-based plasticity can also be written recurrently.

\subsection*{Local plasticity rules}
The above energy gradients (\cref{eq:dWvr,eq:dWqr}) lead to the gradient-based plasticity of the corresponding synaptic weights from a presynaptic neuron $j$ to a postsynaptic neuron $i$,
\begin{align}
  \dot{W}_{V,ij}^h&\propto\; b^h_{t,i} \sum_k q^h_{t,k} S^h_{V,t,kj} \;, \label{eq:dotwV}  \\  
  \dot{W}_{K,ij}^h&\propto\; q^h_{t,i} \sum_k b^h_{t,k}S^h_{K,t,kj} \label{eq:dotwK} \;,\\
  \dot{W}_{Q,ij}^h&\propto\; x_{t,j} \sum_k b^h_{t,k} S^h_{Q,t,ik} 
  \label{eq:dotwQ} 
 \;.
\end{align}
The first two plasticity rules are of the form of a postsynaptic quantity (indexed by $i$) times a sum that depends on the presynaptic neuron (indexed by $j$). 
In the spirit of dendritic gain modulation, it is conceivable that such learning rules are neuronally implemented.
A term in the sum, e.g.~the product $q^h_{t,k} S^h_{V,t,kj}$ in eq.~\ref{eq:dotwV}, may be represented by a pyramidal neuron (indexed by $kj$) with recurrent basal input producing $S^h_{V,t,kj}$ (see eq.~\ref{eq:Sv}) and gain modulated by the apical input $q^h_{t,k}$. The output of these pyramidal neurons are summed up across $k$ in a dendritic branchlet (indexed by $ij$) of the postsynaptic neuron $i$, that is itself modulated by the postsynaptic dendritic signal $b^h_{t,i}$. The same branchlet $ij$ hosts the synapse $W_{V,ij}^h$ from the presynaptic neuron $x_{t,j}$ to the postsynaptic neuron $v_{t,j}^h$ that is modified according to \cref{eq:dotwV}. 

The third plasticity rule (\cref{eq:dotwV}) has a straightforward implementation. It is driven by the correlation between pre- and postsynaptic activity. The presynaptic activity is $x_{t,j}$, and the postsynaptic (dendritic) activity is represented by a sum of synaptic inputs from presynaptic pyramidal neurons indexed by $ki$ and representing the products $b^h_{t,k} S^h_{Q,t,ik}$. These products are formed by recurrent basal inputs to the pyramidal neurons producing $S^h_{Q,t,ik}$ (see \cref{eq:Sq}) and gain modulated by the apical error input $b^h_{t,k}$.

\subsection*{Pseudocodes annotated with the proposed biological substrates}

\begin{algorithm}
\caption{Multihead linear self-attention in cortico-thalamic circuits --- Learning}\label{alg:overview}
\begin{algorithmic}

\State \# Inference, see \cref{alg:inference}
\State output $\gets$ 0
\For{each head}
  \State Update KV memory following \cref{eq:M} \Comment{Layer 2/3 --- thalamo-cortical}
  \State Increment output following \cref{eq:y} \Comment{Layer 5 --- cortico-thalamic}
\EndFor
\State
\State \# Learning
\State error $\gets$ next token - output
\For{each head}\Comment{Gradient-based synaptic plasticity}
  \State Increment slow synaptic variables following \cref{eq:Sv,eq:Sk,eq:Sq}  \Comment{Case $\phi(x)=x$}
  \State Update weights by gradient descent following \cref{eq:dWo,eq:dWvr,eq:dWkr,eq:dWqr} 
\EndFor
\end{algorithmic}
\end{algorithm}

\begin{algorithm}
\caption{Multihead linear Self-Attention in cortico-thalamic circuits --- Inference}\label{alg:inference}
\begin{algorithmic}
\Require $\bm x \in \mathbb{R}^{d_e}$, $\bm M^h\in \mathbb{R}^{d_v\times d_k}$, $\bm W_V^h\in \mathbb{R}^{d_v\times d_e}$, $\bm W_K^h\in \mathbb{R}^{d_k\times d_e}$, $\bm W_Q^h\in \mathbb{R}^{d_k\times d_e}$, $\bm W_O^h\in \mathbb{R}^{d_e \times d_v}$
\State $\bm y \gets \texttt{zeros}(d_e)$
\For{$h = 1,\dots,n_h$}
\State $\bm v^h \gets \bm W_V^h \bm x$ \Comment{Core thalamo-cortical --- Layer 2/3 basal}
\State $\bm k^h \gets \bm W_K^h \bm x$ \Comment{Matrix thalamo-cortical --- Layer 2/3 apical}
\State $\bm q^h \gets \bm W_Q^h \bm x$ \Comment{Matrix thalamo-cortical --- Layer 5 basal}
  \For{$i = 1,\dots,d_v$}
    \For{$j = 1,\dots,d_k$}
      \State $\bm M^h[i,j] \gets \bm M^h[i,j] + \bm v^h[i] \cdot \phi(\bm k^h[j])$ \Comment{Recurrent integration --- Layer 2/3 soma}
      \State $\bm Z^h[i,j] \gets \bm M^h[i,j] \cdot \phi(\bm q^h[j])$ \Comment{Layer 2/3 to Layer 5 apical --- Layer 5 soma}
      \For{$k = 1,\dots,d_e$}
        \State $\bm y[k] \gets \bm y[k] + \bm W_O^h[k,i] \cdot \bm Z^h[i,j]$ \Comment{cortico-thalamic --- Higher-order nucleus}
      \EndFor
    \EndFor
  \EndFor
\EndFor
\State\Return $\bm y$
\end{algorithmic}
\end{algorithm}

\begin{algorithm}
\caption{An equivalent formulation with a single memory variable for both inference and plasticity}\label{alg:single}
\begin{algorithmic}
\Require $\bm x \in \mathbb{R}^{d_e}$, $\bm x_f \in \mathbb{R}^{d_e}$, $\bm M^h\in \mathbb{R}^{d_e\times d_e}$, $\bm W_V^h\in \mathbb{R}^{d_v\times d_e}$, $\bm W_K^h\in \mathbb{R}^{d_k\times d_e}$, $\bm W_Q^h\in \mathbb{R}^{d_k\times d_e}$, $\bm W_O^h\in \mathbb{R}^{d_e \times d_v}$
\State $\bm y \gets \bm 0$
\For{$h = 1,\dots,n_h$}
\State $\bm M^h \gets \bm M^h + \bm x \bm x^{\mathsf{T}}$
\State $\bm y \gets \bm y + \bm W_O^h\bm W_V^h\bm M^h\bm W_K^{h\mathsf{T}}\bm W_Q^h \bm x$
\EndFor
\State $\bm e \, \gets \bm x_f - \bm y$
\For{$h = 1,\dots,n_h$}
\State $\bm W_O^h \gets \bm W_O^h + \lambda \left[-\bm W_O^h + \bm e \bm x^\mathsf{T}\bm W_Q^{h\mathsf{T}}\bm W_K^{h}\bm M^{h\mathsf{T}}\bm W_V^{h\mathsf{T}}\right]$
\State $\bm W_V^h \gets \bm W_V^h + \lambda \left[-\bm W_V^h + \bm W_O^{h\mathsf{T}}\bm e \, \bm x^\mathsf{T}\bm W_Q^{h\mathsf{T}}\bm W_K^h\bm M^h\right]$
\State $\bm W_K^h \gets \bm W_K^h + \lambda \left[-\bm W_K^h + \bm W_Q^{h}\bm x \, \bm e^\mathsf{T}\bm W_O^{h}\bm W_V^h\bm M^h\right]$
\State $\bm W_Q^h \gets \bm W_Q^h + \lambda \left[-\bm W_Q^h + \bm W_K^h \bm M^h \bm W_V^{h\mathsf{T}}\bm W_O^{h\mathsf{T}}\bm e \, \bm x^\mathsf{T}\right]$
\EndFor
\State\Return $\bm y$
\end{algorithmic}
\end{algorithm}

\newpage
\section*{Acknowledgments}
We thank Timothée Proix for discussions on human speech processing and intracortical recordings, Nicolas Deperrois for implementing a previous version of our cortical transformers, Katharina Wilmes and Ausra Saudargiene for working out a time-continuous version within a different project, and Michael Marmaduke Woodman for discussing how to implement it in The Virtual Brain. This work was funded by the European Union's Horizon Europe Programme under the Specific Grant Agreement No. 101147319 (EBRAINS 2.0 Project). We declare no competing interests.

\printbibliography

\end{document}